\documentclass[journal=jacsat,manuscript=article]{achemso}

\usepackage[version=3]{mhchem} 
\usepackage{graphicx}
\usepackage{dcolumn}
\usepackage{bm}
\usepackage{siunitx}
\usepackage{amsmath}
\usepackage[dvipsnames]{xcolor}
\usepackage[version=3]{mhchem} 
\usepackage{siunitx}  

\usepackage{comment}
\usepackage{multirow}
\usepackage{array}
\usepackage{textpos}
\usepackage{rotating} 
\usepackage{esvect}

\usepackage{graphicx}
\usepackage{dcolumn}
\usepackage{bm}
\usepackage{float}

\usepackage[utf8]{inputenc}
\usepackage{float}
\usepackage[T1]{fontenc}
\usepackage{mathptmx}
\usepackage{siunitx}  
\usepackage{amsmath}
\usepackage[hidelinks]{hyperref}
\usepackage[all]{hypcap}
\usepackage[noabbrev, capitalise]{cleveref}
\usepackage{amssymb}

\usepackage{natbib}

\author{Hanqing Liu}
\email{liuhanqing@nudt.edu.cn}
\affiliation[Delft University of Technology]
{College of Electronic Science and Technology, National University of Defense Technology, Deya Road 109, 410073 Changsha, China}
\alsoaffiliation[Delft University of Technology]
{Department of Precision and Microsystems Engineering, Delft University of Technology, Lorentzweg 1, 2628 CD Delft, The Netherlands}
\author{Saurabh Lodha}
\affiliation[Indian Institute of Technology Bombay]
{Department of Electrical Engineering, Indian Institute of Technology Bombay, 400076, India}
\author{Herre S. J. van der Zant}
\affiliation[Delft University of Technology]
{Kavli Institute of Nanoscience, Delft University of Technology, 2628 CJ Delft, The Netherlands}
\author{Peter G. Steeneken}
\affiliation[Delft University of Technology]
{Department of Precision and Microsystems Engineering, Delft University of Technology, Lorentzweg 1, 2628 CD Delft, The Netherlands}
\alsoaffiliation[Delft University of Technology]
{Kavli Institute of Nanoscience, Delft University of Technology, 2628 CJ Delft, The Netherlands}
\author{Gerard J. Verbiest}
\email{G.J.Verbiest@tudelft.nl}
\affiliation[Delft University of Technology]
{Department of Precision and Microsystems Engineering, Delft University of Technology, Lorentzweg 1, 2628 CD Delft, The Netherlands}

\title
  {Optomechanical method for characterizing thermal transport across van der Waals interfaces }

\abbreviations{IR,NMR,UV}
\keywords{American Chemical Society, \LaTeX}

\begin{document}


\begin{abstract}
 For the development of nanoscale electronics and photonics using atomically thin two-dimensional (2D) materials, it is important to realize van der Waals (vdW) interfaces with low thermal resistance, to minimize performance reduction caused by heat accumulation. However, characterizing the thermal interface resistance between vdW materials is still a challenge. Here, we introduce a novel optomechanical methodology to characterize the thermal transport across interfaces in 2D heterostructures. We first determine the specific heat and thermal conductivity as the function of temperature for the upper and lower material layers separately and then extract the thermal boundary conductance (TBC) of the heterostructure from its thermal time constant. We obtain a TBC of $2.41 \pm 1.03$ and $4.14 \pm 1.74$~\si{MW m^{2} K^{-1}} for FePS$_3$/WSe$_2$ and MoS$_2$/FePS$_3$ interfaces, respectively, which are comparable to values reported in the literature. Moreover, they agree with a Debye model including the acoustic impedance mismatch of flexural phonons. This work enables efficient thermal management down to the nanoscale and offers new insights into energy dissipation in vdW heterostructures.
\end{abstract}

\textbf{Keywords:} {2D interface, thermal boundary conductance, optomechanical measurement}
\section{Introduction}

The material class of two-dimensional (2D) materials comprises a wide range of mechanically flexible and atomically thin compounds for the next-generation electronic, photonic and thermoelectric applications \cite{ban2023emerging,steeneken2021dynamics,balandin2011thermal}. For creating unique designer materials or devices that address these applications, stacking 2D materials is needed to tailor the desired functionality \cite{zhang2023van}. However, to ensure heat can be transported away efficiently, low thermal resistance at the interfaces between the 2D materials is important to prevent temperature rise and concomitant performance degradation \cite{ong2019energy}. The heat generated in stacked van der Waals (vdW) devices, i.e. 2D heterostructures, travels both in-plane through each 2D material layer, the substrate, and out-of-plane direction across vdW interfaces including 2D/2D and 2D/substrate interfaces. These heat transport pathways are mainly characterized by the thermal conductivity of the materials and the thermal boundary conductance (TBC) on the interfaces, respectively \cite{vaziri2019ultrahigh,huang2025enhancement}. As a result of the weak vdW bonding between vdW interfaces, large thermal isolation can arise which dominates the heat flow in the system \cite{kim2021extremely}. Therefore, characterizing the thermal transport across vdW interfaces is critical for optimizing the thermal management and thereby enable a pathway towards high performance 2D heterostructures.


So far, a variety of experimental techniques have been developed to determine the TBC at vdW interfaces, with transient electrical \cite{li2017thermal,koh2010heat,buckley2021anomalous} and steady-state Raman thermometry \cite{yalon2017temperature,liu2022energy,yalon2017energy} emerging as the most widely adopted approaches. However, both methods pose significant challenges: electrical thermometry requires complicated device fabrication and is affected by thermal contact resistance effects, while Raman thermometry typically suffers from limited temperature resolution, resulting in substantial measurement uncertainties. These methodological constraints have compromised the precision of thermal transport characterization across vdW interfaces, contributing to the considerable variation in the reported TBC values among existing studies. For example, the TBC values for the graphene/SiO$_2$ interface measured by electrical thermometry vary from 83 to 178~\si{MW/m^2/K} at room temperature \cite{chen2009thermal}, and are much higher than the values below 35~\si{MW/m^2/K} typically obtained by Raman thermometry \cite{vaziri2019ultrahigh,yang2014thermal}.
 
In this Letter, we demonstrate an optomechanical non-contact method for characterizing thermal transport across vdW interfaces. The proposed methodology allows us to simultaneously extract the thermal expansion coefficient, the specific heat and the in-plane thermal conductivity of different type of 2D materials, and finally determine the TBC across vdW interface. It involves determining the cool down time of a suspended membrane that is heated by a power-modulated laser and measuring its deflection, that is driven by thermal expansion with a second laser \cite{liu2024optomechanical,liu2023tuning}. Accordingly, both the temperature-dependent mechanical resonance frequency of the membrane and characteristic thermal time constant of the membrane are measured. A major advantage of the method is that no physical contact needs to be made to the 2D heterostructure, such that its pristine properties are probed and no complex device fabrication is needed. Our results on the TBC of FePS$_3$/WSe$_2$ and MoS$_2$/FePS$_3$ show good agreement with reported values in the literature.   

\section{Fabrication and methodology}

\begin{figure}
	\centering
	\includegraphics[width=1\linewidth,angle=0]{"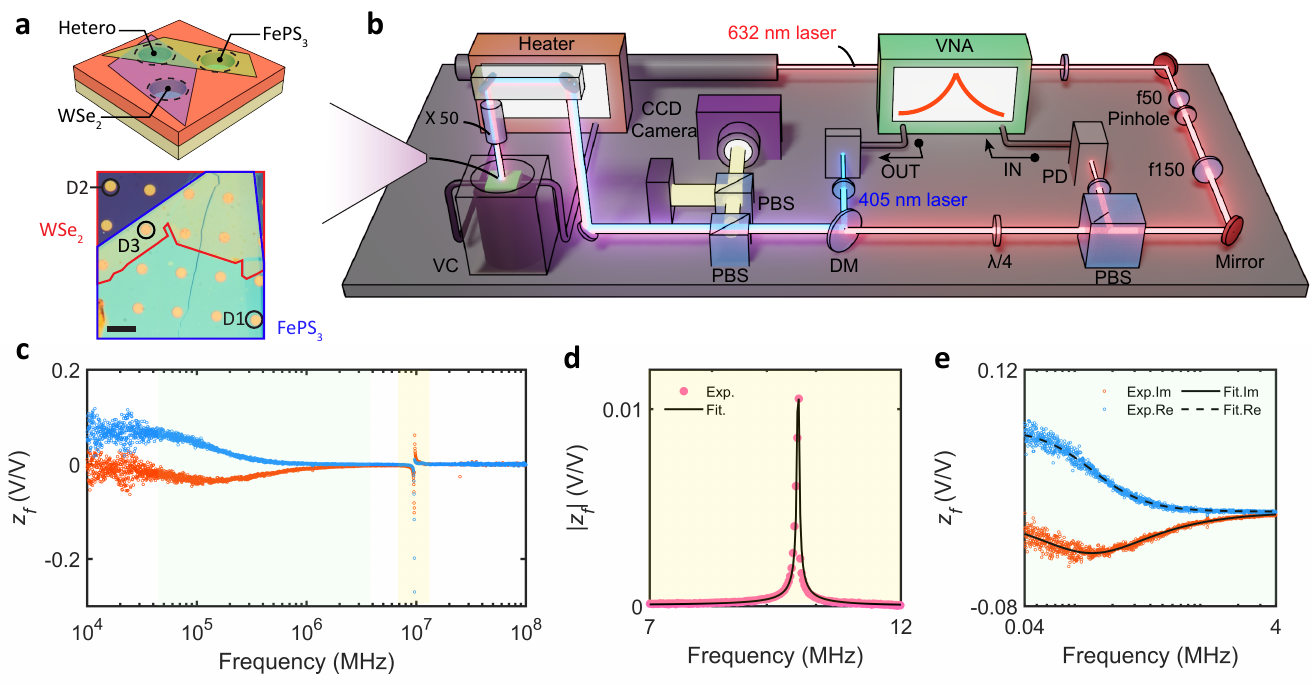"}
	\caption{ Sample characterization and experimental setup. \textbf{a} Schematic (top) and optical image (bottom) of the fabricated 2D resonators. Scale bar: \SI{10}{\micro\meter}. D1, FePS$_3$ device; D2, WSe$_2$ device; D3, FePS$_3$/WSe$_2$ heterostructure device. \textbf{b} Setup of optomechanical measurement. VNA, vector network analyzer; PD, photodiode; PBS, polarized beam splitter; DM, dichroic mirror; VC, vacuum chamber. \textbf{c} Frequency response of device D1, including real (red) and imaginary (blue) parts of the motion $z_f$. \textbf{d} Resonant peak of device D1 measured at MHz regime (points), which is fitted with a harmonic model (drawn line) to extract the resonance frequency $f_0$ of device D1. \textbf{e} Thermal signal measured at kHz regime, including imaginary (red) and real (blue) parts. The imaginary part is fitted with Eq.~\ref{eq:1} (drawn lines) to obtain the thermal time constant $\tau$ of device D1.   }
	\label{fig:a}
\end{figure}

We fabricate 2D nanomechanical resonators by transferring two flakes of different 2D materials on the same chip, such that the flakes partly overlap. Since the chip contains circular cavities in a SiO$_2$ layer on Si, with a depth of 285~\si{nm} and radius of 3~\si{\mu m}, suspended membranes including both of the individual 2D materials and the stacked heterostructure are formed. As illustrated in Fig.\ref{fig:a}a, devices D1 and D2 are made of suspended FePS$_3$ and WSe$_2$ membranes, respectively, while the central overlapping region forms a FePS$_3$(upper)/WSe$_2$(lower) heterostructure, device D3. To determine the Young's modulus $E$ of each membrane, we use the AFM to indent the center of the suspended area with a force while measuring the cantilever indentation \cite{castellanos2012elastic}. As listed in Table~\ref{table_1}, we extract $E=98.3$~\si{GPa} and $93.1$~\si{GPa} for devices D1 and D2, respectively, which are in good agreement with typical values found in literature \cite{kumar2015thermoelectric, vsivskins2020magnetic}. In addition, we fabricate another heterostructure resonator device made of MoS$_2$(upper)/FePS$_3$(lower) flakes, and further experimentally test it with the same methodology. All details about the thickness and Young's modulus measurement of devices D1-D3, as well as the MoS$_2$/FePS$_3$ device can be found in SI section 1. 

The setup for the optomechanical measurements \cite{Siskins2021tunable, liu2024enhanced}, is shown in Fig.~\ref{fig:a}b. A power-modulated blue diode laser ($\lambda=405~$\si{nm}) photothermally actuates the resonator, while a He-Ne laser ($\lambda=632$~\si{nm}), of which the reflected laser power depends on the position of the membrane, is used to detect the motion of the resonator. The power-modulation of the blue laser is supplied by a Vector Network Analyzer (VNA), which also analyzes the photodiode signal containing the reflected laser power and converts that to the response amplitude, $|z_f|$, of the resonator in the frequency domain (see Fig.~\ref{fig:a}c). All measurements were done in vacuum at a pressure below 10$^{-5}~\si{mbar}$. In Fig.~\ref{fig:a}d, $|z_f|$ shows a clear fundamental resonance peak in MHz frequencies, which can be fitted by a harmonic oscillator model given by $|z_{f}| = \frac{A_{\text{res}}f_0^2}{Q\sqrt{(f_0^2-f^2)^2+(f_0f/Q)^2}}$, where $f_0$ is the fundamental resonance frequency, $f$ is the driving frequency, $A_{\text{res}}$ is the vibration amplitude at resonance and $Q$ is the quality factor. For device D1, we obtain a $f_0=9.53$~\si{MHz} and $Q=160$. Moreover, we also find a maximum in the imaginary part of $z_f$ at kHz frequencies (see Fig.~\ref{fig:a}e), which is attributed to the thermal expansion of the membrane that is time-delayed with respect to the modulated blue laser power, because it takes a time $\tau$ for the temperature of the membrane to rise \cite{dolleman2018transient}. The signal at driving frequencies $f$ far below the resonance frequency $f_0$ can be expressed as:
\begin{equation}
    z_f = \frac{A_{\text{th}}}{i2\pi f \tau+1},
   \label{eq:1}
\end{equation}
where $A_{\text{th}}$ and $\tau$ are the thermal expansion amplitude and thermal time constant of the membrane, respectively. The red and blue laser powers are fixed at 0.9 and 0.13~\si{mW} respectively, which are low enough to ensure linear vibration of the resonators. We extract $\tau$ by fitting the measured imaginary part of $z_f$ to Eq.~(\ref{eq:1}) (see Fig.~\ref{fig:a}e). Here, we obtain the maximum of $\text{Im}(z_f)$ at around 366.19~\si{kHz} for device D1, corresponding to $\tau=(2\pi\times366.19$~\si{kHz}$)^{-1}=434.62$~\si{ns}.

\begin{table*}
\caption{\label{table_1}
Characteristics of devices D1 and D2, including radius $R$, thickness $h$, mass density $\rho$, Young's modulus $E$, atomic mass $M$, Poisson ratio $\nu$, Gr$\ddot{\text{u}}$neisen parameter $\gamma$. The values of $\rho$, $M$, $\nu$ and $\gamma$ are taken from literature \cite{kargar2020phonon,kumar2015thermoelectric, vsivskins2020magnetic}.}

\begin{tabular}{lccccccc}
  \hline\hline
 & $R$ (\si{\mu m}) & $h$ (\si{nm}) & $\rho$ (\si{kg~m^{-3}}) & $E$ (\si{GPa}) & $M$ (\si{g~mol^{-1}}) & $\nu$ & $\gamma$  \\
\hline
D1 (FePS$_3$) & 3 & 34.5 & 3375 & 93.1 & 183 & 0.304 & 1.80
\\
D2 (WSe$_2$) & 3 & 8.6 & 9320 & 98.3 & 342 & 0.19 & 0.79 \\

  \hline\hline
\end{tabular}
\end{table*}

\section{Results}

\begin{figure}
	\centering
	\includegraphics[width=1\linewidth,angle=0]{"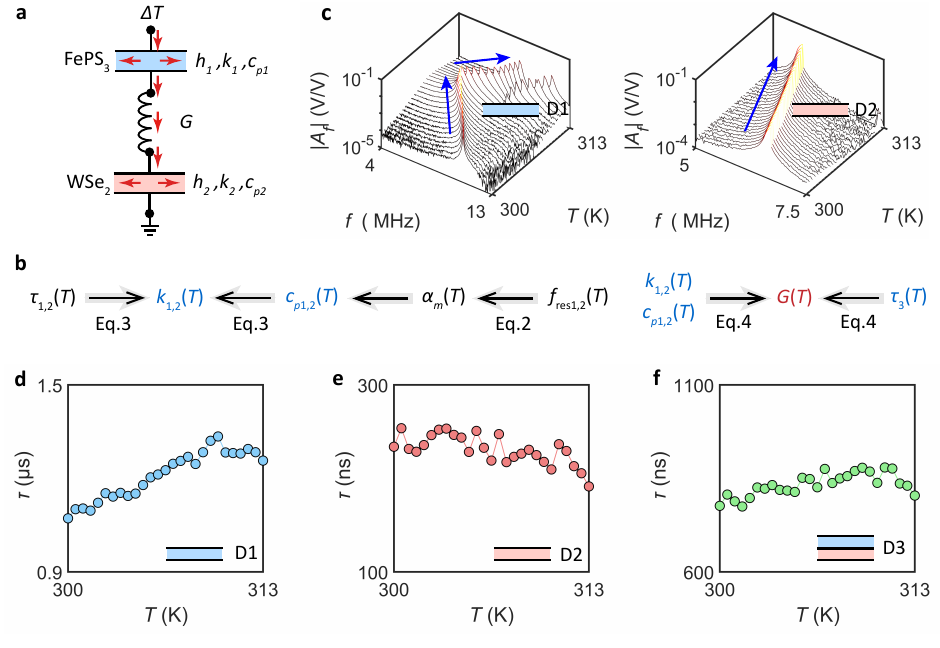"}
	\caption{Temperature-dependent optomechanical measurements on devices D1-D3. \textbf{a} Equivalent model of heat propagation on FePS$_3$/WSe$_2$ membrane. \textbf{b} The proposed procedure to determine the values of TBC $G$. \textbf{c} Resonant peak measured as the function of temperature $T$ for devices D1 and D2. \textbf{d-f} Thermal time constant $\tau$ measured as the function of $T$ for devices D1-D3, respectively. }
	\label{fig:b}
\end{figure}
\begin{figure} 
	\centering
	\includegraphics[width=1\linewidth,angle=0]{"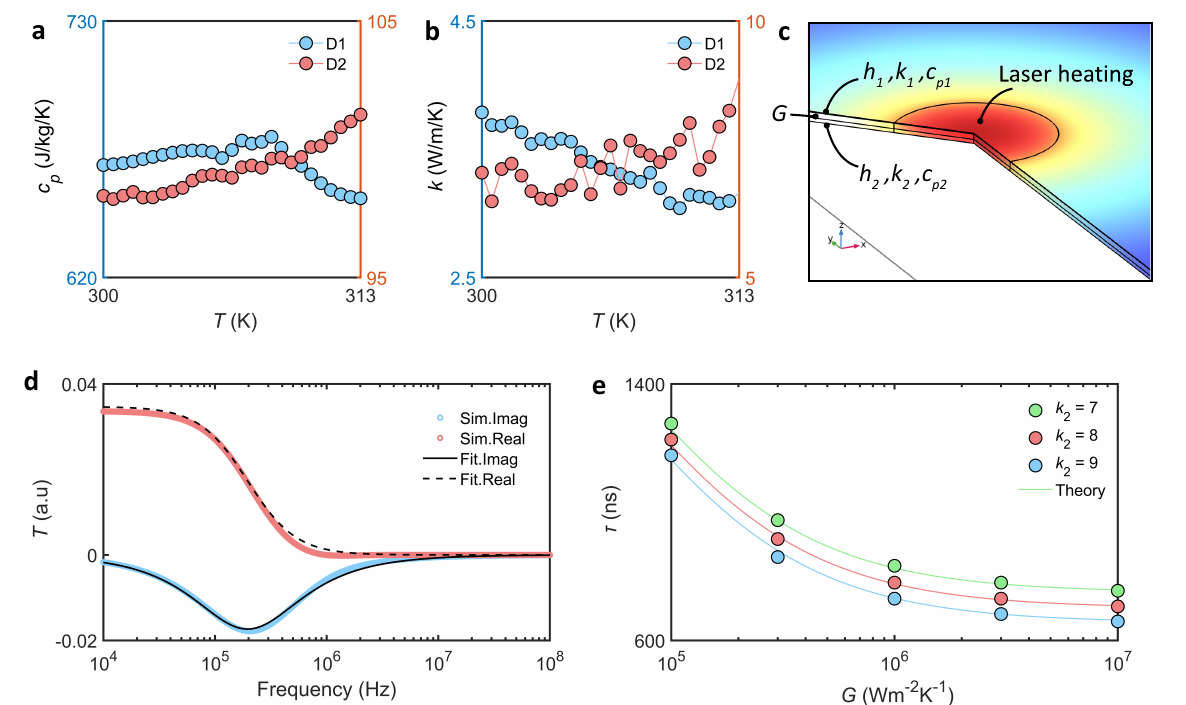"}
	\caption{Characterizing the thermal properties of 2D resonators. \textbf{a} Specific heat $c_p$ and \textbf{b} thermal conductivity $k$ as the function of temperature $T$ for devices D1 and D2, respectively. \textbf{c} Schematic diagram of 2D heterostructure resonator in COMSOL simulation software. \textbf{d} Temperature distribution of the membrane versus heating rate. Points, simulation results; drawn lines, fitting by Eq.~\ref{eq:1}. \textbf{e} Estimations of thermal time constant $\tau_3$ versus $G$ for device D3. Points, simulation results; drawn lines, calculation by Eq.~\ref{eq: tau3 versus G}.   
 }
	\label{fig:c}
\end{figure}

Figure.~\ref{fig:b}a shows the equivalent model that describes the heat transport in the heterostructure resonator. To determine the TBC $G$, the thermal properties of pure FePS$_3$ and WSe$_2$ membranes, including the specific heat $c_p$ and thermal conductivity $k$, need to be determined in advance. We start by heating up the fabricated devices D1 and D2. When changing the temperature, the thermal expansion coefficient (TEC) $\alpha_m$ of the membrane, which is higher than that of the silicon substrate $\alpha_{\text{Si}}$, changes the strain in the membrane. This results in a resonance frequency change of 2D nanomechanical resonators, which can be used for probing the thermal properties \cite{ye2018electrothermally,zhang2020coupling}. As shown in Fig.~\ref{fig:b}c, we observe a decrease of $f_0$ with increasing $T$ for device D2, which is in agreement with trends shown in literature \cite{wang2021thermal} and can be attributed to a reduction in tension when the material thermally expands. Following our previous study \cite{liu2024optomechanical}, the relation between $f_0$ and the thermally induced in-plane displacement $U$ for a clamped circular plate is given by:
\begin{equation}
\frac{\text{d}f_0^2}{\text{d}T } = c_t\frac{\text{d}U}{\text{d}T }, 
   \label{eq:TEC calculation}
\end{equation}
where $\frac{\text{d}U}{\text{d}T } = - R[\alpha_m(T)-\alpha_{\text{Si}}(T)]$, $\nu$ is the Poisson ratio, $d=2R$ is the diameter of the membrane, and $c_t = \frac{13.34E}{\pi^2d^2\rho(1-\nu)}$. The flow chart (left) depicted in Fig.~\ref{fig:b}b exhibits how optomechanical measurements enable a precise pathway for studying the thermal properties of heterostructure resonator. Firstly, we extract the TEC $\alpha_m(T)$ of the membrane from the obtained $\frac{\text{d}U}{\text{d}T}$ using Eq.~\ref{eq:TEC calculation}, where the values of $\alpha_\text{Si}(T)$ are taken from literature \cite{okada1984precise}. Next, the specific heat $c_p$ of the membrane is extracted directly from $\alpha_m$ using the thermodynamic relation $c_p=  3\alpha_m K / \gamma\rho$ \cite{vsivskins2020magnetic}, where $K=\frac{E}{3(1-2\nu)}$ is the bulk modulus, and $\gamma$ is the Gr$\ddot{\text{u}}$neisen parameter of the membrane taken from literature (see Table~\ref{table_1}). As shown in Fig.~\ref{fig:c}a, the obtained $c_{p2}(T)$ for device D2 is about 100$\pm$2~\si{J kg^{-1} K^{-1}}, which is comparable to the literature values \cite{kumar2015thermoelectric}. Then, by solving the heat equation in the membrane with an appropriate initial temperature distribution and well-defined boundary conditions, the thermal conductivity of the membrane can be extracted from the measured $\tau$ and the obtained $c_p$ through:
\begin{equation}
    \tau_j = \frac{R^2c_{pj}\rho_j}{\mu_j^2 k_j }, j=1,2 
   \label{eq: k versus T}
\end{equation}
where $k_j$ is the thermal conductivity of 2D materials and the in-plane diffusive eigenvalue for single-layer circular plate is $\mu_j^2=5$ \cite{liu2024optomechanical}. Hence, by substituting the obtained $c_{p2}$ and the measured $\tau_2$ (Fig.~\ref{fig:b}e) into Eq.~\ref{eq: k versus T}, we extract $k_2 = 3.4\pm0.3$~\si{W m^{-1} K^{-1}} for device D2 (see Fig.~\ref{fig:c}b). In the following we will explain how to extract the TBC from the measured thermal time constant of the heterostructure.

For device D1, we observe an initial decrease in $f_0$ with increasing $T$ towards a minimum frequency, followed by a continuous increase. This is attributed to the thermally-induced buckling of mechanical resonators, causing by a loaded compression since $\alpha_m > \alpha_{\text{Si}}$. In this case, a factor $\delta$ connected with the deflection of the membrane is added to Eq.~\ref{eq:TEC calculation} in order to accurately determine the relation between $f_0$ and $U$ under buckling \cite{liu2024enhanced}. The detailed expression of $\delta$ can be found in SI section 1. We further extract $c_{p1}$= 665$\pm$11~\si{J kg^{-1} K^{-1}} and $k_1 = 7.6\pm1.2$~\si{W m^{-1} K^{-1}} for device D1, as plotted in Figs.~\ref{fig:c}a and \ref{fig:c}b, respectively, which are comparable to the literature values \cite{kargar2020phonon}.

\begin{figure}
	\centering
	\includegraphics[width=0.8\linewidth,angle=0]{"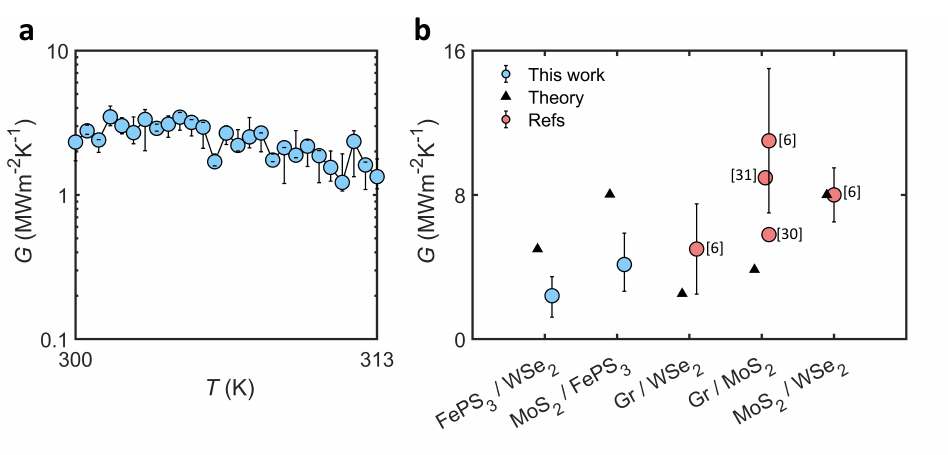"}
	\caption{Quantifying TBC $G$ in 2D heterostructure resonators. \textbf{a} Dependence of $G$ on temperature $T$ for device D3 from 300 to 313~\si{K}. \textbf{b} Comparison of the extracted $G$ in this work with the reported values from literature. Black triangles, Theoretical estimation of $G$ by the built Debye model.}
	\label{fig:d}
\end{figure}

After characterizing the thermal properties of FePS$_3$ and WSe$_2$ membrane separately, we move on to the estimation of TBC $G$ in heterostructure device D3. To figure out the dependence of $\tau_3$ on $G$, we employed a 3D heat transport model in COMSOL, as illustrated in Fig.~\ref{fig:c}c. The structural parameters $R$, $h_{1,2}$, $\rho_{1,2}$ are used as given for FeP$_3$ and WSe$_2$ membranes. The laser pulse with harmonic perturbation irradiates at the center of the top surface, and the spot size is fixed as its realistic value $R_0 = 0.5$~\si{\mu m}. The boundary conditions on the interface are set as $k_1\frac{\partial T_1}{\partial z} =k_2\frac{\partial T_2}{\partial z} = -G(T_1-T_2)$, where $T_1$ and $T_2$ are the temperature distribution of FePS$_3$ and WSe$_2$ membrane, respectively. Using $c_{p1} = 690$~\si{J/kg/K}, $c_{p2} = 90$~\si{J/kg/K}, $k_{1} = 3.5$~\si{W/m/K}, $k_{2} = 8$~\si{W/m/K} and $G = 10$~\si{MW/m^2/K} in the model, we simulate the normalized temperature distribution of the heterostructure as plotted in Fig.~\ref{fig:c}d. A thermal time constant $\tau_3$ of 779.5~ns can be extracted by fitting the imaginary part of $T$ with Eq.~\ref{eq:1}. We then obtain a negative dependence of $\tau_3$ on $G$ by the simulation, as shown in Fig.~\ref{fig:c}e (solid points), indicating that a higher TBC on the vdW interface will accelerate the overall heat transfer of the system. To shed light on the physical mechanism of $\tau_3$ versus $G$, we solve the Fourier heat conduction equations in cylindrical coordinates for a bi-layered circular plate with similar boundary conditions as in the COMSOL simulation (see SI section 2 for derivation), which allows to obtain the expression:
\begin{equation}
    \tau_3 = \tau_{3,int} + \frac{c_{p2}\rho_2}{k_2\left( \eta^2 + \gamma^2 \right)},
    \label{eq: tau3 versus G}
\end{equation}
where the eigenvalues are
\begin{equation*}
    \eta^2 = \left(\frac{2.4048}{R}\right)^2  \quad \text{and} \quad  \gamma^2  = \frac{Gk_1}{k_1k_2(h_2-h_1) + G(k_2h_1h_2-k_2h_1^2 - k_1h_1h_2 ) }. 
\end{equation*}
The first part in Eq.~\ref{eq: tau3 versus G} represents the intrinsic thermal time constant of the  heterostructure device when $G \rightarrow + \infty$. Considering the case of two-layer 2D materials connected in parallel, we have $\tau_{3,int} = \frac{R^2(c_{p1}h_1\rho_1+ c_{p2}h_2\rho_2)}{\mu_3^2(h_1k_1 + h_2k_2)}$, where the in-plane diffusive eigenvalue $\mu_3^2$ for heterostructure is determined as 5.36 by COMSOL simulation (see SI section 3); the second part represents the contribution from $G$-dominated thermal diffusivity, which is obtained from the solution of Fourier heat conduction equations. As shown in Fig.~\ref{fig:c}e, we obtain a good agreement of $\tau_3$ versus $G$ between the estimation and simulation under different $k_2$. As a result, through substituting the extracted $c_{p1,2}$, $k_{1,2}$ and the measured $\tau_3$ (see Fig.~\ref{fig:b}f) into Eq.~\ref{eq: tau3 versus G}, we finally extract $G$ about $2.41 \pm 1.03$~\si{MW m^{-2} K^{-1}} for device D3 in the range from 300 to 313~\si{K}, as shown in Fig.~\ref{fig:d}a. Using the same approach, we also determine $G=3.91 \pm 1.78$~\si{MW m^{2} K^{-1}} for the fabricated MoS$_2$/FePS$_3$ heterostructure resonator (see SI section 1 for details). 

The obtained $G$ for our fabricated devices are comparable to the reported values for vdW interfaces made of MoS$_2$, WSe$_2$ or graphene \cite{vaziri2019ultrahigh,liu2015thermal,wu2020interfacial}, as shown in Fig.~\ref{fig:d}b (points). To better understand the physical mechanism of TBC, we follow the literature \cite{vaziri2019ultrahigh,little1959transport} and use a Debye model wherein $G$ is proportional to the product of the overlapping phonon density of states (PDOS) areas of the two materials, the mass density mismatch, and the derivative of the Bose-Einstein distribution function with respect to temperature (see SI section 4 for details). Here, we only consider the PDOS for flexural (ZA) phonons, which are known to dominate the out-of-plane heat conduction in 2D heterostructure devices \cite{liu2023tuning,lindsay2010flexural}. As plotted in Fig.~\ref{fig:d}b (drawn line), the theoretical estimation of $G$ by the Debye model matches well with our experimental results as well as the other reported values from literature. We conclude that the characteristics of ZA phonons in FePS$_3$ attribute to a small area of PDOS overlap at FePS$_3$-based vdW interface, resulting in the low values of TBC for our FePS$_3$/WSe$_2$ and MoS$_2$/FePS$_3$ devices in this work.

\section{Discussion and conclusions}

Compared to other methods for measuring the thermal boundary conductance of vdW interfaces, our proposed approach offers several advantages, as outlined in Table~\ref{tab:method comparison}. For instance, Raman thermometry requires large temperature changes to detect the frequency shifts in Raman mode, necessitating measurements over a broad temperature range to precisely determine the slope $\chi_T$ of the peak shift versus temperature. In the case of MoS$_2$, for example, $\chi_T$ is only ~\si{cm^{-1}/K}, which means given a typical resolution limit of 0.25~\si{cm^{-1}} for a Raman microscope, a temperature rise of at least 20~\si{K} is needed to achieve reliable results \cite{sahoo2013temperature}. In addition, the crosstalk of Raman modes in different 2D materials, e.g., MoS$_2$ and WS$_2$, will significantly affect the resolution of $\chi_T$. For electrical-thermometry, either thick crystals or stiff 2D materials like graphene have to survive the complicated fabrication procedures including lithography and etching \cite{wang2017thermal}. In contrast, for the presented contactless optomechanical method, one only needs to suspend membranes over cavities in a SiO$_2$/Si substrate, while the accuracy can be maintained within 0.1~\si{K} by using a sensitive heater. Hence, our proposed methodology provides a precise and applicable platform for characterizing the thermal transport in vdW interfaces.

Considering a detection range of thermal signal from 10~\si{kHz} to 1~\si{MHz} on VNA, our proposed methodology is applicable to 2D resonators that can be excited photothermally with thermal time constants $\tau$ between 160~\si{ns} and 16~\si{\mu s}. The 2D resonator should not be perfectly flat. In reality, inhomogeneities due to uneven adhesion or residues on the vdW interface could lead to multiple smaller corrugations and wrinkles superimposed in the membrane during thermal expansion, which could result in estimation errors when extracting specific heat from Eq.~\ref{eq:TEC calculation}. Exfoliated 2D flakes are transferred onto the substrate using a dry transfer method, causing polymer residues on the vdW interfaces that could affect the measured TBC. As a result, we primarily selected the samples with clean surfaces for optomechanical measurement under optical microscope (see Fig.~\ref{fig:a}a). In addition, Eqs.~\ref{eq: k versus T} and \ref{eq: tau3 versus G} are not applicable for monolayer and bilayer graphene, in which the scattering of flexural phonons at the boundary between the supported and suspended parts of the 2D material dominates the value of the thermal time constant in measurement \cite{liu2023tuning,dolleman2017optomechanics}. 

It should be noted that $\mu_3^2$ in Eq.~\ref{eq: tau3 versus G} is related to the thickness and thermal properties of 2D materials, as well as the laser spot size. We verify that within the measured temperature range, the offset of $\mu_3^2$ is below 0.2~\% compared to the fixed constant 5.36 (see SI section 3).Moreover, our proposed methodology is more efficient when $G<$10~\si{MW m^{2} K^{-1}}, where the slope of $\tau$ versus $G$ is large enough to ensure a high accuracy of obtaining $G$ by the measured $\tau$ (see Fig.~\ref{fig:c}e). Finally, although the measured results of TBC in this work are well captured by Debye model as shown in Fig.~\ref{fig:d}b, we note that the flexural phonon behaviors in 2D materials are highly related to their dispersion relation, surface tension, crystal grain size, and temperature \cite{kuang2016thermal}. These factors significantly affect the thermal transport in 2D heterostructure and should be further studied using the presented optomechanical methodology, which would help us to better understand the phonon scattering mechanisms in vdW interfaces and why the theoretical estimation of $G$ by the Debye model can explain the experimental observations surprisingly well.

\begin{table*}
\caption{\label{tab:method comparison}
Comparison of different methods for measuring the TBC of vdW interfaces.} 
\begin{tabular}{llll}
  \hline\hline
 & Raman thermometry & Electrical thermometry & Optomechanics \\
\hline
Required temperature range & 50$-$100~\si{K} \cite{liu2017thermal, ye2023ultra} &  10$-$50~\si{K} \cite{li2017thermal}  & $<10$~\si{K}
\\
Sample preparation &  Easy  & Difficult  &  Easy
\\
Applicability to 2D materials &  Applicable & Limited  & Applicable 
\\
  \hline\hline
\end{tabular}
\end{table*}

In conclusion, we demonstrated an optomechanical approach for probing the thermal transport across vdW interfaces in heterostructure resonators made of few-layer FePS$_3$/WSe$_2$ and MoS$_2$/FePS$_3$. By measuring the resonance frequency and thermal time constant of the fabricated devices as a function of temperature, we separately extract the specific heat and in-plane thermal conductivity of the upper and lower layers, and finally extract the thermal boundary conductance $G$ on the vdW interface. The obtained values of $G$ are in good agreement with values reported in the literature. Compared to other methods, the presented contactless optomechanical approach requires a smaller temperature range, allows for easy sample fabrication, and is applicable to any 2D materials. This work not only advances the fundamental understanding of phonon mismatch in vdW interfaces, but potentially also enables studies into the use of strain engineering and heterostructures for optimizing the heat flow in 2D nanodevices.

\begin{acknowledgement}

P.G.S. and G.J.V. acknowledge support by the Dutch 4TU federation for the Plantenna project. H.S.J.v.d.Z. and P.G.S. acknowledge funding from the European Union’s Horizon 2020 research and innovation program under grant agreement no. 881603. H.L. acknowledges the financial support from China Scholarship Council and the National Natural Science Foundation of China (No. 62401592).

\end{acknowledgement}




\bibliography{achemso-demo}

\newpage

\setcounter{equation}{0}
\setcounter{figure}{0}
\setcounter{table}{0}
\setcounter{page}{1}
\makeatletter
\renewcommand{\theequation}{S\arabic{equation}}
\renewcommand{\thefigure}{S\arabic{figure}}
\renewcommand{\thetable}{S\arabic{table}}

\section{SI section 1: Mechanical characterization of the fabricated devices}

\begin{figure}
	\centering
    
	\includegraphics[width=1\linewidth,angle=0]{"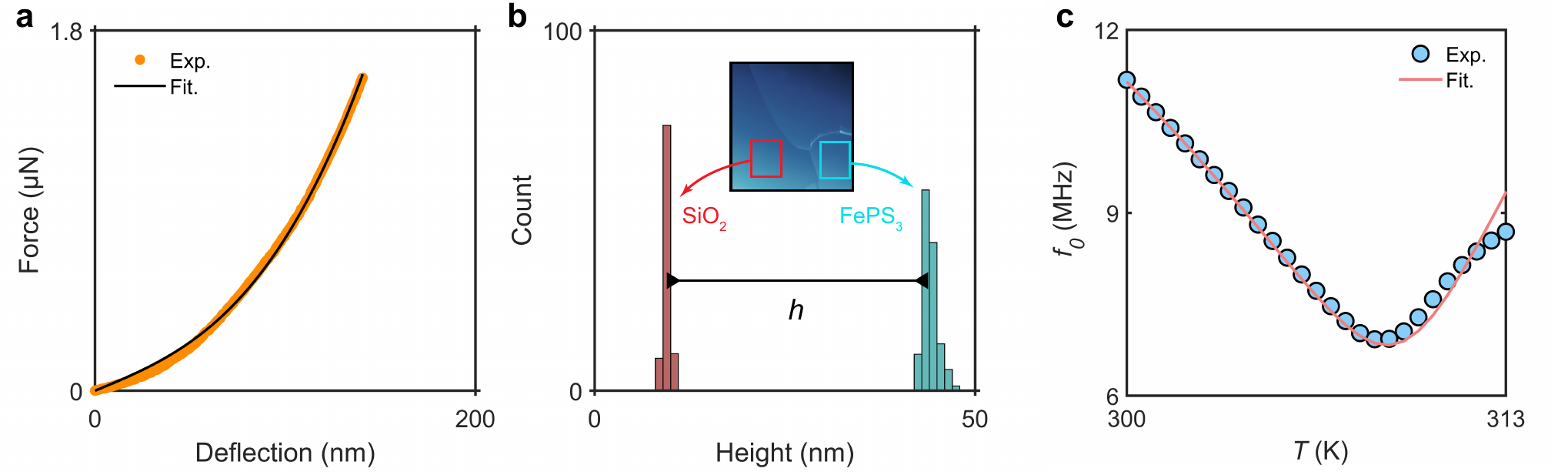"}
	\caption{ AFM measurements on FePS$_3$ membrane (device D1). \textbf{a} AFM indentation results for device D1 (orange points), where the Young’s modulus of the membrane is extracted by fitting the measured force to the cantilever deflection (black line). \textbf{b} Heigh histogram of the substrate (red), as well as FePS$_3$ membrane (cyan), measured by AFM. Insert, AFM scanning image on the boundary of FePS$_3$ flake. \textbf{c} Points, measured resonance frequency versus temperature for device D1; drawn line, estimation by mechanical buckling model.}
	\label{fig:SI_a}
\end{figure}

Force-indentation AFM method is a reliable mechanical probe of 2D material membrane defection in response to a force applied at its center by an AFM cantilever tip. We first image the fabricated membrane using AFM tapping/AC mode and thus to align the tip exactly above the center of the membrane; then the membrane is indented with a preset value of maximum force to obtain its corresponding deflection behavior. We use the AFM cantilever with typical stiffness $k_c=10$~\si{N/m} in the force indentation. As shown in Fig.~\ref{fig:SI_a}a, the measured force-deflection curve can be used to extract the pretension $n_0$ and Young's modulus $E$ of the FePS$_3$ membrane, using the expression \cite{castellanos2012elastic}:
\begin{equation}
   F = ( \frac{16\pi D}{r^2}\Delta ) + n_0\pi\Delta+Etq^3(\frac{\Delta^3}{r^2}),
   \label{eq:methodology_indentation}
\end{equation}
where $\Delta$ is the deflection, $D = Eh^3/(12(1-\nu^2))$ is the bending rigidity of the membrane which can be ignored if $h$ is small enough, $\nu$ is Poisson ratio, $n_0=Eh\epsilon_0/(1-\nu)$ is the pretension, $\epsilon_0$ is the built-in strain, and $q = 1/(1.05-0.15\nu-0.16\nu^2)$ is the geometry-related parameter. By fitting Eq.~\ref{eq:methodology_indentation} with the measured curve, we thus extract $E = 93.1$~\si{GPa} for device D1, with $R=3$~\si{\mu m} and $h=34.5$~\si{nm}.

AFM tapping mode is typically used to determine the thickness $h$ of 2D membranes. However, since 2D membrane is bumpy with bulges and collapses, it is not fairly accurate to extract the realistic thickness from only a few scanning profiles. Therefore, we count the height of the entire region which contains thousands of points, as plotted in Fig.~\ref{fig:SI_a}b. The mean height of both substrate and FePS$_3$ membrane are thus calculated by the statistics, so as to obtain the mean thickness of the membrane about 34.5~\si{nm}. 

For device D1, we observe a first decrease and then increase $f_0$ as temperature $T$ increases, which is attributed to a thermally-induced buckling of the membrane from the boundary loaded compression. According to our previous study \cite{liu2024optomechanical}, a factor $\delta$ need to be induced into the $T$-dependent $f_0$ relation, thus Eq.~2 is modified as:
\begin{equation}
    \frac{\text{d}f_0^2}{\text{d}T } =  c_t\left(1-\frac{32}{\frac{16}{3(1-\nu^2)}\frac{z_{free}h^2}{z^3} + 10.7} \right) \frac{\text{d}U}{\text{d}T },
   \label{eq:case II}
\end{equation}
where $U$ is the thermally changed in-plane displacement from boundary, $z$ is the central deflection of the plate and $z_{free}$ is the central deflection of the plate at free state when $U=0$ (without loading). We first extract $z_{free} = 4.3$~\si{nm} and the function of $z$ on $T$ from the optomechanical measurement, and thus obtain a good agreement of $f_0$ versus $T$ between estimation and experiment. Then we substitute $z_{free}$ and $z(T)$ into Eq.~\ref{eq:case II} to extract $\frac{\text{d}U}{\text{d}T}$, which allows us to obtain the thermal expansion coefficient $\alpha_{m1}$, the specific heat $c_{p1}$ and the thermal conductivity $k_1$ of device D1 in turns.

\begin{figure}
	\includegraphics[width=1\linewidth,angle=0]{"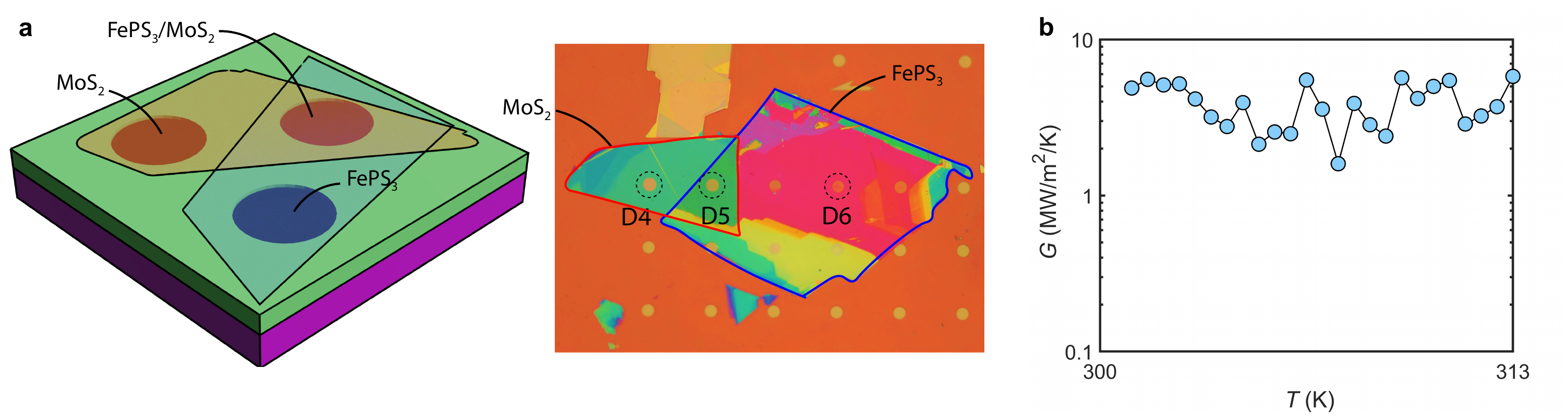"}
	\caption{ Devices D4-D6 made of MoS$_2$/FePS$_3$ membranes. \textbf{a} Structure diagram and optical image; \textbf{b} Extracted TBC $G$ as the function of temperature $T$ for device D5. }
	\label{fig:SI_2}
\end{figure}

Besides the FePS$_3$/WSe$_2$ devices D1-D3 in the main text, we also fabricated another MoS$_2$/FePS$_3$ devices D4-D6, as shown in Fig.~\ref{fig:SI_2}a. Through AFM measurements we extract $h_4$ = 10.1~\si{nm}, $h_5=113.2$~\si{nm}, $E_4=103$~\si{GPa}, $E_5=300$~\si{GPa}. Using the proposed methodology, we finally obtain the TBC $G=3.91 \pm 1.78$~\si{MW m^{2} K^{-1}} for the MoS$_2$/FePS$_3$ interface in device D5, as shown in Fig.~\ref{fig:SI_2}b.

\section{SI section 2: Fourier heat conduction equations for bi-layered circular plate}

\begin{figure}
	\centering
    \captionsetup{justification=centering}
	\includegraphics[width=0.7\linewidth,angle=0]{"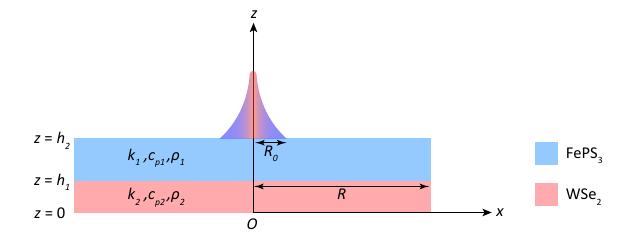"}
	\caption{ Illustration of double-layer circular plate.}
	\label{fig:TIC_model_1}
\end{figure}

In this section, we calculate the thermal transport in a bi-layered circular plate, so as to analysis the role that the TBC $G$ of 2D heterostructure plays on its thermal time constant $\tau_h$. For a bi-layered circular plate, as depicted in Fig.~\ref{fig:TIC_model_1}, the homogeneous of heat conduction equations in cylindrical coordinates can be expressed as:
\begin{equation}
    \begin{cases}
    \frac{\partial T_1}{\partial t} = \kappa_1\left ( \frac{\partial^2T_1}{\partial x^2} + \frac{1}{R} \frac{\partial T_1}{\partial x} + \frac{\partial^2 T_1}{\partial z^2} \right), \quad \text{in} \quad 0<z<h_1 \quad 0<x<R, \\
     \frac{\partial T_2}{\partial t} = \kappa_2\left ( \frac{\partial^2T_2}{\partial x^2} + \frac{1}{R} \frac{\partial T_2}{\partial x} + \frac{\partial^2 T_2}{\partial z^2} \right), \quad \text{in} \quad h_1<z<h_2 \quad 0<x<R, 
    \end{cases}
   \label{eqs.heat equations hetero} 
\end{equation}
where the heights of top and bottom layer are at $h_1$ and $h_2$, $\kappa_1 = \frac{k_1}{c_{p_1} \rho_1}$ and $\kappa_2 = \frac{k_2}{c_{p_2} \rho_2}$ are the thermal diffusivity, $k$, $c_p$ and $\rho$ are thermal conductivity, specific heat and mass density of the membranes, respectively. To obtain temperature distributions $T_1$ and $T_2$, we combine Eqs.~\ref{eqs.heat equations hetero} with the boundary and initial conditions, given as:
\begin{equation}
    \begin{cases}
    T_1 = 0, \quad \text{at} \quad x=R, \\
    T_1 = 0, \quad \text{at} \quad z=0, \\
    T_2 = 0, \quad \text{at} \quad x=R, \\
    -k_1\frac{\partial T_1}{\partial z} = G(T_1-T_2), \quad \text{at} \quad z=h_1 \\
    k_1\frac{\partial T_1}{\partial z} = k_2\frac{\partial T_2}{\partial z} , \quad \text{at} \quad z=h_1, \\
    \frac{\partial T_2}{\partial z} = 0, \quad \text{at} \quad z=h_2.
    \end{cases}
   \label{eqs.heat conditions hetero} 
\end{equation} 
Accordingly, using the separation of variables, $T_1$ and $T_2$ can be expressed as:
\begin{equation}
    \begin{cases}
    T_1(x,z,t) = \psi(x) Z_1(z) \Gamma(t),  \\
    T_2(x,z,t) = \psi(x) Z_2(z) \Gamma(t).  \\
    \end{cases}
   \label{eqs.heat hetero general} 
\end{equation} 
Combining the boundary and initial conditions in Eq.~\ref{eqs.heat conditions hetero}, the general solutions for Eq.~\ref{eqs.heat hetero general} are:
\begin{equation}
    \begin{cases}
    T_1(x,z,t) = {\sum_{m=1}^\infty } {\sum_{n=1}^\infty } A_{mn} J_0(\eta_mx) \sin({\xi_n z}) e^{-\lambda_{mn}^2t},  \\
    T_2(x,z,t) = {\sum_{m=1}^\infty } {\sum_{n=1}^\infty } B_{mn} J_0(\eta_mx) [\cos({\gamma_n z}) +  \tan({\gamma_n z_2})\sin({\gamma_n z})] e^{-\lambda_{mn}^2t},  \\
    \end{cases}
   \label{eqs.heat conditions hetero} 
\end{equation} 
where $m=1,2,3,...$ and $n = 1,2,3,...$ are the $m$-th and $n$-th eigenvalues values, $A_{mn}$ and $B_{mn}$ are integration parameters. $\xi_n$ and $\gamma_n$ can be determined by the relations:

\begin{equation}
\lambda_{mn}^2 = \kappa_1(\eta_m^2 + \xi_n^2) = \kappa_2(\eta_m^2 + \gamma_n^2)
   \label{eqs.hetero condition 1} 
\end{equation} 
and
\begin{equation}
\begin{vmatrix}
    k_1\xi_n\cos(\xi_nh_1)+ G\sin(\xi_nh_1) & -G\cos({\gamma_n h_1}) & -G\sin(\gamma_nz_1) \\ 
     k_1\xi_n\cos(\xi_nh_1) & k_2\gamma_n \sin(\gamma_nh_1) & -k_2\gamma_n \cos(\gamma_nh_1)\\
     0 & \sin(\gamma_nh_2) & -\cos(\gamma_nh_2)
\end{vmatrix} = 0.
   \label{eqs.hetero condition} 
\end{equation} 
As a result, the thermal time constant $\tau$ of the heterostructure, dominated by $\lambda_{mn}$, is related to the thermal properties of the membranes, as well as their TBC $G$.

Let us now focus on the solution of Eqs.~\ref{eqs.hetero condition 1} and ~\ref{eqs.hetero condition}. Using Taylor expansion of $\sin{x} = x$ and $\cos{x} = 1$ as $x\rightarrow 0$, Eq.~\ref{eqs.hetero condition} is simplified to:
\begin{equation}
\begin{vmatrix}
    k_1\xi_1+ G\xi_1 h_1 & -G & -G\gamma_1h_1 \\ 
     k_1\xi_1 & k_2\gamma_1 \gamma_1h_1 & -k_2\gamma_1 \\
     0 & \gamma_1h_2 & -1
\end{vmatrix} = 0.
   \label{eqs.hetero condition simplified} 
\end{equation} 
Combine Eqs.~\ref{eqs.hetero condition 1} and ~\ref{eqs.hetero condition simplified}, we final extract:
\begin{equation}
   \mu_1 = \sqrt{\frac{Gk_1}{k_1k_2(h_2-h_1)+G(k_2h_1h_2-k_2h_1^2-k_1h_1h_2)}}
   \label{eqs.solution of gamma} 
\end{equation} 
As a result, Eq.~\ref{eqs.solution of gamma} is used to express the in-plane diffusive eigenvalue $\mu_3^2$ for heterostructure in the Eq.~4 in the main text.

\section{SI section 3: COMSOL simulation on 2D heterostrcuture}

\begin{figure}
	\centering
	\includegraphics[width=1\linewidth,angle=0]{"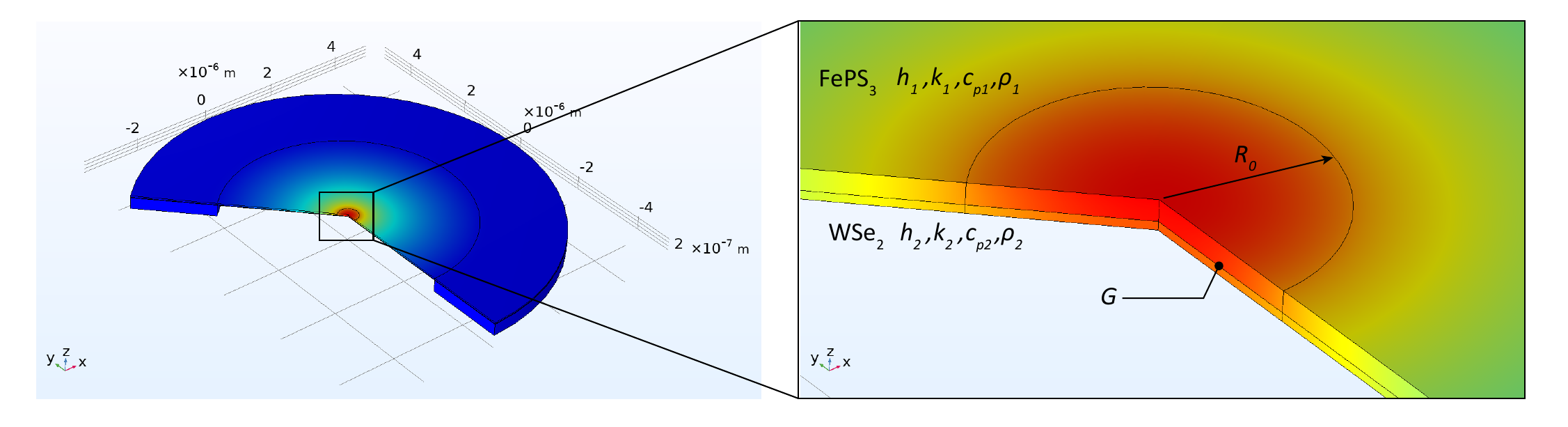"}
	\caption{COMSOL model for simulating thermal transport in a double-layer circular membrane.}
	\label{fig:TIC_COMSOL_model}
\end{figure}

Although the temperature distribution $T_1$ and $T_2$ for top and bottom layers can be solved with the form of Green's functions as reported in literature, it is complicated to obtain the analytical solution for $\lambda_{mn}^2$, much less considering the size of laser spot $R_0$ is smaller than $r$ and the existing of Si substrate in realistic case. Therefore, we build the COMSOL simulation model to directly estimate the relation between $\tau$ and $G$ in the heterostructure device. As shown in Fig.~\ref{fig:TIC_COMSOL_model}, we first set the structural parameters as $R=3$~\si{\mu m}, $R_0 = 0.5$~\si{\mu m}, $h_1 = 34.5$~\si{nm}, $h_2 = 8.6$~\si{nm}, $\rho_1 = 3375$~\si{kg m^{-3}} and $\rho_2 = 9320$~\si{kg m^{-3}} as listed in Table~1 in the main text. Then by defining the initial and boundary conditions of thermal transport, the thermal parameters $k_{1,2}$, $c_{p1,2}$ and $G$, as well as a harmonic perturbation of heating with a frequency varies from 10~kHz to 100~MHz, we obtain the average temperature of the membrane as a function of heating frequency, as plotted in Fig.~3d in the main text. 

In the main text, we determine the in-plane diffusive eigenvalue $\mu_3^2$ for heterostructure circular plate is about 5.36, and thus obtain a good agreement of $\tau$ versus $G$ between estimation by Eq.~4 and simulation. In fact, $\mu_3^2$ is also highly dependent on the thermal properties of 2D materials, including including $k_{1,2}$ and $c_{p1,2}$. To further investigate it, we simulate the thermal transport when $G\rightarrow\infty$, under case (i): $k_1=6.5$~\si{W/m/K}, $k_2=3$~\si{W/m/K}, $c_{p1}=100$~\si{J/kg/K} and $c_{p2}=680$~\si{J/kg/K} (slow transport); and case (ii) $k_1=8.8$~\si{W/m/K}, $k_2=3.8$~\si{W/m/K}, $c_{p1}=98$~\si{J/kg/K} and $c_{p2}=650$~\si{J/kg/K} (fast transport). The ranges of $k_1$, $k_2$, $c_{p1}$ and $c_{p2}$ above are fixed by the extracted maximum and minimum values of our sample as shown in Figs.~3a and 3b. According to the simulation results, we extract $\mu_3^2$ is equal to 5.354 and 5.371 for cases (i) and (ii), respectively, which means the fitted $\mu_3^2$ has an offset lower than 0.2~\% compared to its realistic values.

\section{SI section 4: Debye model of flexrual phonons mismatch}
For 2D materials, the TBC across the interface is primarily dominated by the out-of-plane or flexural phonon (ZA) modes. The flexural phonon density of states (PDOS) is given by $C_{qz}  = 1/(4\pi\sqrt{\sigma/\rho h_{s}})$, where $\sigma$ is the flexural rigidity of the 2D material, $\rho$ is the mass density and $h_{s}$ is the single layer thickness of 2D materials. To calculate the PDOS overlap between upper and lower materials, we also need the cut-off energies $\theta_{zd}$, which is also called as Deybe energies, for the participating phonon branches. According to the literature, we have $\theta_{zd} = \kappa q_{zd}^2$, where $\kappa = \sqrt{\sigma/\rho h_{s}}$ and $q_{zd} = \sqrt{8\pi/\sqrt{3}h_{s}^2}$.

\begin{figure}
	\centering
	\includegraphics[width=1\linewidth,angle=0]{"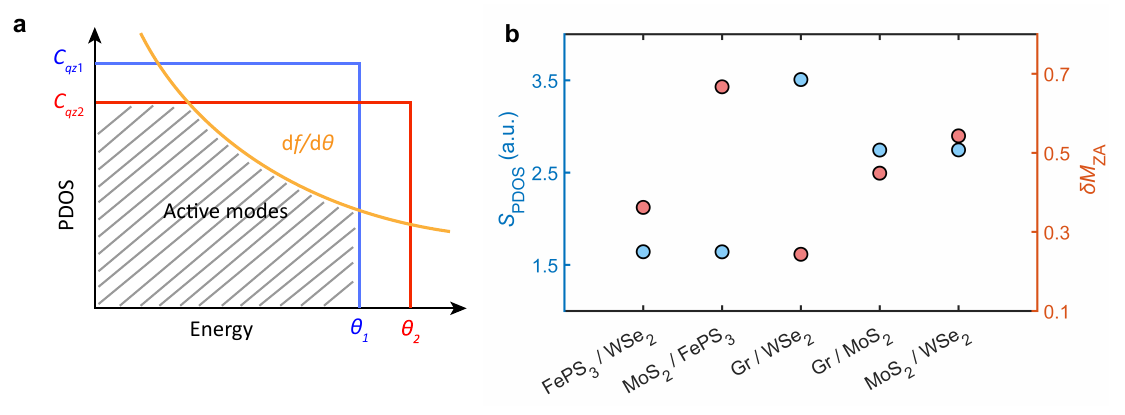"}
	\caption{Calculation of the built Debye model for 2D vdW interfaces. \textbf{a} PDOS overlap area. \textbf{b} The extracted PDOS overlap $S_{\text{PDOS}}$ (left $y$-axis) and density mismatch $\delta M_{\text{ZA}}$ (right $y$-axis) for different types of interfaces. }
	\label{fig:last}
\end{figure}

Following literature, we use a Debye model to evaluate the TBC across 2D vdW interface. We start from estimating the PDOS overlap area $S_{\text{PDOS}}$ between the upper and lower 2D materials, as shown in Fig.~\ref{fig:last}a (shaded area). In this case, $S_{\text{PDOS}}$ is expressed by:
\begin{equation}
   S_{\text{PDOS}} = \int_0^{\text{min}(\theta_{zd_1},\theta_{zd_1})} \cdot \text{min}(C_{qz_1},C_{qz_2})\cdot \frac{\text{d}f(\theta)}{\text{d}T} \text{d}\theta,
   \label{eqs.solution of S} 
\end{equation} 
where $k_b$ is Boltzmann constant, $T$ is the temperature and $f(\theta)$ is the Bose-Einstein distribution function given by $f(\theta)=\frac{1}{e^{{\theta}/{k_bT}}-1}$. The $C_{qz}$ and $\theta_{zd}$ for FePS$_3$, MoS$_2$, WSe$_2$ and graphene are extracted as listed in Table~\ref{tab:debye}. These values are subsitituted into Eq.~\ref{eqs.solution of S}, which allows us to extract the $S_{\text{PDOS}}$ of different types of 2D vdW interfaces, as shown in Fig.Fig.~\ref{fig:last}b.

\begin{table*}
\caption{\label{tab:debye}
The PDOS $C_{qz}$ and cut-off energies $\theta_{zd}$ of 2D materials.} 
\begin{tabular}{lllll}
  \hline\hline
 & FePS$_3$ & MoS$_2$ & WSe$_2$ & graphene \\
\hline
$C_{qz}$ (\si{m^{-2}s}) & 0.68$\times 10^{5}$ &  1.13$\times 10^{5}$ &  1.45$\times 10^{5}$  & 1.65$\times 10^{5}$
\\
$\theta_{zd}$ (meV) &  26.6 &  17.8  & 12.4  &  41
\\
  \hline\hline
\end{tabular}
\end{table*}

After determing the PDOS overlap area, we then focus on the transmission of flexural phonons across the 2D vdW interfaces. This can be simply evaluated by an acoustic mismatch model, since the vdW interfaces are nearly perfect and any surface asperities (e.g.
dangling bonds) are much smaller than the phonon wavelengths \cite{vaziri2019ultrahigh}. As a result, the transmission is given by the density mismatch between upper/lower materials:
\begin{equation}
   \delta M_{\text{ZA}}=\text{min}(\rho_1/\rho_2,\rho_2/\rho_1).
   \label{eqs.solution of mismatch} 
\end{equation} 
Finally, the TBC is proportional to the product of the PDOS overlap and the density mismatch:
\begin{equation}
   G \propto S_{\text{PDOS}} \cdot \delta M_{\text{ZA}}.
   \label{eqs.solution of TBC} 
\end{equation} 

Using Eqs.~\ref{eqs.solution of S} to \ref{eqs.solution of TBC}, we normalize the product $S_{\text{PDOS}} \cdot \delta M_{\text{ZA}}$ of device D3 (FePS$_3$/WSe$_2$) and set it as 1, thus to calibrate all the other types of interfaces. As shown in Fig.~\ref{fig:last}b, the measured low values of TBC for our fabricated FePS$_3$/WSe$_2$ and MoS$_2$/FePS$_3$ interfaces can be explained by the small overlapped area of ZA phonons.

\end{document}